\documentclass[%
 reprint,
groupedaddress,
 amsmath,amssymb,
 aps,
 pra,
]{revtex4-1}
\usepackage[T1]{fontenc}
\usepackage[final]{graphicx}
\usepackage{dcolumn}
\usepackage{amsfonts}
\usepackage{amssymb}
\usepackage{pbox}
\usepackage{amsmath}
\usepackage{amsthm}
\usepackage{mathtools}
\usepackage[normalem]{ulem}

\DeclareMathOperator{\Tr}{Tr}
\DeclareMathOperator{\diag}{diag}
\DeclareMathOperator{\const}{const}

\DeclareMathOperator{\re}{re}
\DeclareMathOperator{\im}{im}

\usepackage{caption}
\usepackage{subcaption}
\usepackage{mathtools}
\usepackage{braket}
\renewcommand\bra[1]{{\langle{#1}|}}
\renewcommand\ket[1]{{|{#1}\rangle}}

\usepackage{chngcntr}
\usepackage{multirow}
\usepackage{tabularx}
\usepackage[labelsep=period]{caption}
\date{}
\usepackage{graphicx}
\usepackage{enumerate}
\usepackage{calc}
\usepackage{indentfirst}
\usepackage{booktabs}
\usepackage{dsfont}
\usepackage{arydshln}
\usepackage{bm}
\usepackage{amsthm}
\usepackage{commath}
\usepackage{xcolor}

\usepackage{natbib}
\setcitestyle{square,numbers}
\usepackage[colorlinks]{hyperref}
\usepackage{changepage}

\begin{document}

\title{Dissipative evolution of quantum Gaussian states}
\author{Tomasz Linowski$^{1}$\footnote{linowski@cft.edu.pl}, Alexander Teretenkov$^{2}$, {\L}ukasz Rudnicki$^{1,3}$}

\affiliation{
    $^1$International Centre for Theory of Quantum Technologies, University of Gdansk, 80-308 Gda{\'n}sk, Poland \\
    $^2$Department of Mathematical Methods for Quantum Technologies, Steklov Mathematical Institute of Russian Academy of Sciences, 119991 Moscow, Russia \\
    $^3$Center for Theoretical Physics, Polish Academy of Sciences, 02-668 Warszawa, Poland
}

\date{\today}

\begin{abstract}
Recent works on quantum resource theories of non-Gaussianity, which are based upon the type of tools available in contemporary experimental settings, put Gaussian states and their convex combinations on equal footing. Motivated by this, in this article, we derive a new model of dissipative time evolution based on unitary Lindblad operators which, while does not preserve the set of Gaussian states, preserves the set of their convex combinations, i.e. so-called quantum Gaussian states. As we demonstrate, the considered evolution proves useful both as a description for random scattering and as a tool in dissipator engineering.
\end{abstract}


\maketitle
\section{Introduction}

One of the most prominent families of states in continuous variable quantum mechanics consists of Gaussian states, i.e. states with Gaussian (normal) characteristic functions. Due to their relative simplicity in both analytical description and practical implementation, they found extensive use in fields as varied as quantum optics, information and thermodynamics, among others \cite{gaussian_optics,Gaussian_thermodynamics_Singh_2019,gaussian_information_1,gaussian_information_2}.

Consequently, much interest was devoted to time evolution which preserves the set of Gaussian states. Such evolution is described by the Gorini–Kossakowski–Lindblad-Sudarshan (GKLS) equation generated by a polynomial of at most second degree in the quadrature operators. Over the years, it proved to be successful in studies of quantum thermodynamics \cite{GTOs}, optics \cite{Gaussian_optics_Olivares_2012}, entanglement \cite{using_dissipation_1,Gaussian_entanglement_Isar_2010}, discord \cite{Gaussian_discord_Giorda_2010}, purity \cite{using_dissipation_2,Gaussian_purity_Paris_2003}, fidelity \cite{Gaussian_fidelity_Isar_2008}, steering \cite{Gaussian_steering}, stabilizability \cite{stabilizability_cv_systems,stabilizing_entanglement_in_two_mode_Gaussian_states} and classical limits of quantum mechanics \cite{Gaussian_classical_Graefe_2018,Gaussian_classical_brodier2009markovian}, among others.

Despite the popularity of Gaussian states and dynamics, in comparison, little interest was devoted to the so-called \emph{quantum Gaussian states}, which are a generalization of Gaussian states that also includes their convex combinations \cite{quantum_Gaussianity_Genoni_2013,quantum_Gaussianity_Hughes_2014,quantum_gaussianity_Kuhn_2018,quantum_gaussianity_Lee_2019,quantum_Gaussianity_Walschaers_2021,quantum_gaussianity_Lachman_2022}. However, according to recent developments in quantum resource theories of non-Gaussianity \cite{Gaussianity_resource_Albarelli_2018,Gaussianity_resource_Takagi_2018}, which are motivated by the type of operations available in modern experiments employing continuous variables, Gaussian states and their convex combinations are equally resourceful. From this perspective, the aforementioned restriction to evolution preserving the set of Gaussian states is too severe and should be relaxed to allow the more general quantum Gaussian states.

In this work, we develop an explicit model of time evolution compatible only with this weaker restriction: it preserves the convex hull of Gaussian states and not the Gaussian family of states itself. Despite that, it is fully compatible with the symplectic (covariance matrix) picture used extensively in studies of Gaussian phenomena. The model is derived from the central assumption of unitary Lindblad operators, a class studied first in 1972 within the then-rapidly developing field of quantum dynamical semigroups \cite{unitary_lindbladians_kossakowski_1972,unitary_lindbladians_Kummerer_1987,unitary_lindbladians_Frigerio_1989}. 

The considered evolution has two very different applications depending on the nature of unitary Lindblad operators entering it. For a large number of non-commuting operators, we use a combination of the collision model and kicked top dynamics to show that the evolution describes random scattering, a view consistent with the first findings regarding unitary Lindblad operators \cite{unitary_lindbladians_kossakowski_1972}. On the other hand, for a single Lindblad operator, time evolution may be employed in dissipator engineering, which we demonstrate with an example of entanglement creation in two-mode states.

The article is organized as follows. Section \ref{sec:preliminaries} is devoted to preliminaries: symplectic picture of quantum states and evolution that preserves the set of Gaussian states. In Section \ref{sec:main_result}, we develop the discussed evolution equation and study its basic technical properites: Gaussianity and symplectic representation. In Section \ref{sec:classical_environment}, we consider the evolution as a description of random scattering. In Section \ref{sec:explicit_solutions}, we investigate the stationary solutions of the derived evolution equation, which we then use in Section \ref{sec:engineered_dissipation} in an engineered dissipation scenario for entanglement harvest. We conclude in Section \ref{sec:conclusion}.

\section{Symplectic picture} 
\label{sec:preliminaries} 
Studies of Gaussian states and their evolution often make use of the symplectic picture, which reduces the $N$-mode infinitely-dimensional Hilbert space associated with the density operator to a space of dimension $2N$, which is typically easier to work with. Here, we briefly summarize the relevant information about the symplectic picture, including the so-called covariance matrix and Gaussianity-preserving evolution.

\subsection{Covariance matrix and vector of means}
Let us consider an $N$-mode Hilbert space $\mathcal{H} = \bigotimes_{k=1}^N\mathcal{H}_k$ equipped with $N$ pairs of quadrature operators $\hat{x}_k, \hat{p}_k$, conveniently collected in a single vector
\begin{align} \label{eq:xi}
\hat{\vec{\xi}} \coloneqq \big(\hat{x}_1, \hat{p}_1, \ldots, \hat{x}_N, \hat{p}_N\big)^T.
\end{align}
As the quadrature operators form a basis of operators acting on 
$\mathcal{H}$, every state describing the system can be fully characterized \cite{Ivan_2012} by the complete ($n=1,\ldots,\infty$) set of \emph{$n$-th order correlation functions (correlations)} of the form 
\begin{align}
\braket{\hat{\xi}_{l_1}\ldots\hat{\xi}_{l_n}}
    \coloneqq\Tr\big[\hat{\rho}\,\hat{\xi}_{l_1}\ldots\hat{\xi}_{l_n}\big],
\end{align}
which we also call \emph{$n$-th moments} for short. In many studies, especially those involving Gaussian states, i.e. states with Gaussian characteristic functions \cite{two-mode_gaussian_etc_proper_norm,cv_systems_gaussian_states,gaussian_optics}, it is enough to consider only the first and second moments. The advantage is that, in contrast to the infinitely-dimensional density operator, the first two moments are completely described by a moderate number of degrees of freedom \footnote{More precisely, for an $N$-mode system, there are $2N$ first moments and $4N^2$ second moments. Taking into account the fact that some of the second moments are co-dependent (due to the canonical commutation relations) yields a total of $N(2N+3)$ independent parameters.}.

Information about the first moments is contained in a $2N$-dimensional \emph{vector of means}
\begin{align} \label{eq:vector_of_means}
    \xi_k\coloneqq\braket{\hat{\xi}_k},
\end{align}
while the second moments are encoded in the $2N\times 2N$ \emph{covariance matrix} 
\begin{align} \label{eq:covariance_matrix}
    V_{kk'}\coloneqq\frac{1}{2}
        \braket{\big\{\hat{\xi}_k,\hat{\xi}_{k'}\big\}}-\xi_k\xi_{k'}.
\end{align}
Both $\{\cdot,\cdot\}$ and $[\cdot,\cdot]$ as usual denote commutators and anticommutators respectively.
Any valid covariance matrix has to be positive and fulfill the Heisenberg uncertainty relations (we assume natural units):
\begin{align} \label{eq:uncertainty_principle}
\sqrt{\braket{\hat{x}_k^2}-\braket{\hat{x}_k}^2}
	\sqrt{\braket{\hat{p}_k^2}-\braket{\hat{p}_k}^2}
	\geqslant \frac{1}{2},
\end{align}
where $k\in\{1,\ldots,N\}$, equivalent to \cite{two-mode_gaussian_etc_proper_norm}
\begin{align} \label{eq:Heisenberg_uncertainty_principle}
    V+\frac{i}{2}J \geqslant 0.
\end{align}
Here, $J$ is the \emph{symplectic form}, defined in terms of the canonical commutation relations as
\begin{align} \label{eq:symplectic_form}
    J_{kk'} \coloneqq -i\big[\hat{\xi}_k,\hat{\xi}_{k'}\big],
\end{align}
and explicitly equal to
\begin{align} \label{eq:symplectic_form_explicit}
    J = \bigoplus_{k=1}^N J_2, 
        \quad J_2 \coloneqq
        \begin{bmatrix}
        0&1\\
        -1&0
        \end{bmatrix}.
\end{align}

The symplectic form defines the symplectic group $Sp(2N,\mathbb{R})$ consisting of matrices $K$ of size $2N \times 2N$, such that \footnote{Some define symplectic matrices by an alternative relation $K^T J K = J$. However, both definitions are completely equivalent, since when $K$ is symplectic, so is $K^T$. Indeed, without loss of generality, consider $K J K^T = J$. Then $K^T = J K^{-1} J^{-1}$, which substituted into $K^T J K$ yields $J$. Proof in the other direction is analogous.}
\begin{align} \label{eq:symplectic_matrices_definition}
\begin{split}
    KJK^T=J.
\end{split}
\end{align}  
In this article, special emphasis is put on a subset of symplectic matrices which possess the following exponential representations (both of which are useful depending on the context)
\cite{symplectic_matrices_properties,Lie_algebras_Samelson}
\begin{align} \label{eq:K_exponential_representation}
\begin{split}
    K=e^{JS}\equiv e^{S'J},
\end{split}
\end{align}  
for some symmetric matrices $S$ and $S'=J S J^T$. We stress that while all matrices of the form (\ref{eq:K_exponential_representation}) are symplectic \footnote{This can be seen as follows. By expanding the exponential function into power series, we obtain that $e^{JS}J=Je^{SJ}$. Furthermore, $(e^{JS})^T=e^{-SJ}$ due to the fact that $J^T=-J$. Substituting these into eq. (\ref{eq:symplectic_matrices_definition}) we immediately obtain that $e^{JS}$ is symplectic. The same holds for $e^{S'J}$, since $S'J=JS$.}, not all symplectic matrices are of this form due to the fact that the symplectic group is not compact \cite{symplectic_exceptions_Torre_2005,symplectic_exceptions_Williamson_1939}.


The pair $(V,\vec{\xi})$ defines the \emph{symplectic picture} (also known as the \emph{covariance matrix picture}) of quantum states. All standard notions known from the density operator picture translate in a natural way to the symplectic picture. In particular, just like any density operator can be diagonalized by a unitary operation and is therefore described by its eigenvalues, any covariance matrix can be brought to a diagonal form by a symplectic operation and is described by its symplectic eigenvalues:
\begin{align} \label{eq:symplectic_eigenvalues}
\begin{split}
    1/2\leqslant \nu_1 \leqslant \ldots \leqslant \nu_N.
\end{split}
\end{align}
The symplectic eigenvalues come in pairs, i.e. the diagonalized covariance matrix reads $V_{\diag}=\diag(\nu_1,\nu_1,\ldots,\nu_N,\nu_N)$. Furthermore, they are related to the eigenvalues $\mu_j$ of the matrix $JV$ via
\begin{align} \label{eq:mu_nu}
\begin{split}
     i\mu_{2k} = -i\mu_{2k-1}^* = \nu_k, \quad k\in\{1,\ldots,N\}.
\end{split}
\end{align}

In the case of Gaussian states, the symplectic picture is complete, i.e. it is equivalent to the density operator description. Otherwise, it describes a subset of the system's degrees of freedom.

\subsection{Gaussianity-preserving evolution}
In the theory of quantum dynamical semigroups, the state of the system at time $t\geq 0$ is given by
\begin{align} \label{eq:GKLS_formal_solution}
    \hat{\rho}(t)=e^{t\mathcal{L}\cdot}\hat{\rho}(0).
\end{align}
Here, $\mathcal{L}$ is the \emph{generator of evolution}, which has the general form
\begin{align} \label{eq:GKLS_generator}
    \mathcal{L}\cdot=-i\big[\hat{H},\cdot\big]
        +\sum_{j}\left(\hat{L}_j\cdot\hat{L}_j^\dag
        -\frac{1}{2}\big\{\hat{L}_j^\dag\hat{L}_j,\cdot\big\}\right),
\end{align}
The system Hamiltonian $\hat{H}$ is responsible for unitary evolution, while the Lindblad operators (Lindbladians) $\hat{L}_j$ govern the dissipative part of the dynamics. 

Here and below we use the dot to denote the argument of the generator, e.g. the action $\mathcal{L}\hat{\rho}$ of the generator on a generic state is given by the r.h.s. of eq. (\ref{eq:GKLS_generator}) with the dot replaced by $\hat{\rho}$. On the other hand, the exponential of the generator is to be understood in terms of its repeated application on the state via
\begin{align} \label{eq:GKLS_generator_precise}
    e^{t\mathcal{L}\cdot}\hat{\rho} = \sum_{n=0}^\infty 
        \frac{t^n}{n!} \underbrace{\mathcal{L}\mathcal{L}\ldots\mathcal{L}}_{n \textnormal{ times}} \hat{\rho}.
\end{align}
This convention is followed by us throughout the article.

By differentiating both sides of eq. (\ref{eq:GKLS_formal_solution}) with respect to time, we obtain the \emph{GKLS (Lindblad) equation} \cite{GKS_original,lindblad_original,lindblad_proof_mathematical}:
\begin{align} \label{eq:GKLS}
    \frac{d}{dt}\hat{\rho}=
        -i\big[\hat{H},\hat{\rho}\big]
        +\sum_{j}\left(\hat{L}_j\hat{\rho}\hat{L}_j^\dag
        -\frac{1}{2}\big\{\hat{L}_j^\dag\hat{L}_j,\hat{\rho}\big\}\right).
\end{align}
If the generator is a polynomial of at most second degree in the quadrature operators, the evolution preserves the set of Gaussian states. In such cases, the Hamiltonian equals
\begin{align} \label{eq:hamiltonian_quadratic}
    \hat{H} = \frac{1}{2}\hat{\vec{\xi}}^T G \hat{\vec{\xi}},
\end{align}
where $G$ is a $2N\times 2N$ real, symmetric matrix. The Lindblad operators, on the other hand, equal
\begin{align} \label{eq:dissipation_quadratic}
    \hat{L}_j = \sum_{k=1}^{2N} (\vec{c}_j)_k \hat{\xi}_k,\qquad\vec{c}_j\in\mathbb{C}^{2N},
\end{align}
necessarily being just linear in the quadratures.

Computing the time derivative of the covariance matrix and assuming that the system evolves according to the GKLS equation specified by eqs. (\ref{eq:hamiltonian_quadratic}, \ref{eq:dissipation_quadratic}), we obtain the corresponding equations for the covariance matrix and the vector of means \cite{using_dissipation_1,using_dissipation_2,stabilizability_cv_systems,stabilizing_entanglement_in_two_mode_Gaussian_states}
\begin{align} \label{eq:covariance_evolution_quadratic}
\begin{split}
    \frac{d}{dt}V &= AV + VA^T + J \re C^\dag C J^T,\\
    \frac{d}{dt}\vec{\xi} &= A\vec{\xi},
\end{split}
\end{align}
where $A\coloneqq J\left[G+\im C^\dag C\right]$ and $C_{jk}\coloneqq (\vec{c}_j)_{k}$.

\section{Dissipative evolution stemming from unitary Lindblad operators}
\label{sec:main_result}
Quantum resource theories classify quantum operations and states according to a given physical property, typically corresponding to usefulness with respect to some practical tasks \cite{quantum_resource_theories_Chitambar_2019}. For example, in resource theories of entanglement, entangled states are considered resourceful, while separable states are classified as free \cite{quantum_resource_theory_entanglement_Bennett_1996,quantum_resource_theory_entanglement_HHHH_2009}. Accordingly, operations incapable of creating entangled states from separable ones are also deemed free. Such classification is natural from the experimental point of view, since, like any valuable resource, entanglement is useful yet difficult to obtain, while operations preserving the set of separable states are relatively easy to implement. By calling entanglement a resource, one can better pose and answer practical questions like, e.g., assuming no limits on free operations, how much entanglement is needed to realize a given teleportation protocol? 

In the resource theories of Gaussianity \cite{Gaussianity_resource_Albarelli_2018,Gaussianity_resource_Takagi_2018}, the set of free operations consists of operations routinely available in current experiments employing continuous variable quantum systems. These include Gaussianity-preserving unitary operations, compositions with Gaussian states and homodyne measurements. In such a setting, the emergent free states (which are preserved by the free operations) are \emph{quantum Gaussian}, that is, they consist of Gaussian states and their convex combinations \cite{quantum_gaussianity_Kuhn_2018,quantum_gaussianity_Lee_2019,quantum_gaussianity_Lachman_2022} (we stress that quantum Gaussian states and Gaussian states are not the same, as the former generalize the latter). From this resource-theoretic point of view, it is natural to look for physically meaningful evolution preserving the set of quantum Gaussian states. By definition, such evolution requires no input resources and can be thus used to manipulate a given system at no cost. 

Observe that the usually assumed Gaussian dynamics (\ref{eq:covariance_evolution_quadratic}) already preserve the set of quantum Gaussian states: since they map Gaussian states to Gaussian states, then, by linearity, they also map their convex combinations to other such combinations. Thus, the generator of Gaussian dynamics, given by eq. (\ref{eq:GKLS_generator}) with eqs (\ref{eq:hamiltonian_quadratic}, \ref{eq:dissipation_quadratic}) at the input, preserves the set of quantum Gaussian states. However, in principle, there may exist other generators that preserve the set of quantum Gaussian states without necessarily preserving the set of Gaussian states. This is exactly what we investigate here.

\subsection{The model of time evolution}
Let us go back to the GKLS equation (\ref{eq:GKLS}). Being interested in the dissipative part of the equation only, we can disregard the Hamiltonian term. As for the dissipator, we follow \cite{unitary_lindbladians_kossakowski_1972,unitary_lindbladians_Kummerer_1987,unitary_lindbladians_Frigerio_1989,alicki_reduced_state,Teretenkov2020} and consider a particular case of $M$ Lindblad operators, all being  proportional to unitary operators: 
\begin{align} \label{eq:unitary_Lindblad_operators}
    \hat{L}_j=\sqrt{\gamma_j}\hat{U}_j,
\end{align}
where $\gamma_j\geqslant 0$, $\hat{U}_j\hat{U}_j^\dag=\hat{U}_j^\dag\hat{U}_j=\hat{\mathds{1}}$, and $\hat{U}_j$ is moreover assumed to be Gaussianity-preserving. All our results are based on this central assumption. The corresponding GKLS equation is generated by
\begin{align} \label{eq:unitary_Lindblad_equation_generator}
    \mathcal{L}=\sum_{j=1}^M\gamma_j\left(\hat{U}_j \cdot \hat{U}_j^\dag- \hat{\mathds{1}}\right)
\end{align}
and thus reads
\begin{align} \label{eq:unitary_dissipator}
    \frac{d}{dt}\hat{\rho}=&
        \sum_{j=1}^M\gamma_j\left(\hat{U}_j\hat{\rho}\hat{U}_j^\dag-\hat{\rho}\right).
\end{align}
For convenience, we assume that $\gamma_j$ fulfill $\sum_{j=1}^M\gamma_j=1$. 

We stress that the choice (\ref{eq:unitary_Lindblad_operators}) of Lindblad operators constitutes a certain loss of generality with respect to the general GKLS equation (\ref{eq:GKLS}). For example, in the considered case, operators $\hat{L}_j^\dag\hat{L}_j$ and $\hat{L}_j\hat{L}_j^\dag$ are proportional to the identity, and consequently both commute with any state $\hat{\rho}$. Such property is not fulfilled by generic Lindblad operators.

To see that eq. (\ref{eq:unitary_dissipator}) preserves the set of quantum Gaussian states, we start with a single unitary operation. Since eq. (\ref{eq:unitary_dissipator}) is a subclass of the GKLS evolution, its formal solution is given by eq. (\ref{eq:GKLS_formal_solution}) with generator (\ref{eq:unitary_Lindblad_equation_generator}). For a single Lindbladian, the latter reduces to $\mathcal{L}\cdot=\hat{U}\cdot\hat{U}^\dag-\hat{\mathds{1}}$. The identity commutes with any operator, so 
\begin{align} \label{eq:rho_t_superoperator_trick}
    \hat{\rho}(t)=e^{t \hat{U}\cdot\hat{U}^\dag }e^{-t\hat{\mathds{1}}}\hat{\rho}(0)=
        \sum_{k=0}^\infty p_k(t) \hat{U}^k \hat{\rho}(0) \big(\hat{U}^{\dag}\big)^k,
\end{align}
where 
\begin{align} \label{eq:Poisson}
p_k(t) \coloneqq e^{-t}t^k/k!
\end{align}
is the Poisson distribution.

Similarly, for an arbitrary number of Lindbladians, we have
\begin{align} \label{eq:rho_t_superoperator_trick_general}
    \hat{\rho}(t)=
        \sum_{k=0}^\infty \sum_{l_1\ldots l_k=1}^M p_{l_1\ldots l_k}(t) 
            \hat{U}_{l_k}\ldots\hat{U}_{l_1} \hat{\rho}(0) 
            \hat{U}_{l_1}^\dag\ldots\hat{U}_{l_k}^\dag,
\end{align}
where for $k=0$ the summand is $e^{-t}\hat{\rho}(0)$ and for $k>0$ $p_{l_1\ldots l_k}(t)\coloneqq \gamma_{l_1}\ldots\gamma_{l_k}e^{-t}t^k/k!$.

Any unitary operator has an exponential representation of the form
\begin{align} \label{eq:U_exponential_representation}
\begin{split}
    \hat{U}_j=e^{-i\hat{h}_j},
\end{split}
\end{align}  
for some hermitian operator $\hat{h}_j$, called the operator's \emph{generator} [not to be confused with the generator of the GKLS evolution (\ref{eq:GKLS_generator})]. As is well-known \cite{cv_systems_gaussian_states,Gaussianity_resource_Albarelli_2018,Gaussianity_resource_Takagi_2018}, unitary operations with generators that are polynomials of at most second degree in quadrature operators preserve the set of Gaussian states. Furthermore, if each $\hat{U}_{l_j}$ preserves Gaussian states, then so does $\hat{U}_{l_k}\ldots\hat{U}_{l_1}$, and therefore each of the terms in the sum (\ref{eq:rho_t_superoperator_trick_general}) maps Gaussian states to other Gaussian states.  

Since the sum of Gaussian states is in general not Gaussian, then even for an initial Gaussian state the time-evolved state (\ref{eq:rho_t_superoperator_trick}) is also not Gaussian in general. On the other hand, if the initial state is a convex combination of Gaussian states, then, by linearity, the time-evolved state is also a convex combination of Gaussian states. Therefore, under the assumption that each Lindblad operator (\ref{eq:unitary_Lindblad_operators}) is generated by a polynomial of at most second degree in quadrature operators, eq. (\ref{eq:unitary_dissipator_Gaussian_evolution}) preserves the set of quantum Gaussian states without preserving the set of Gaussian states, as we wanted to show.

\subsection{Representation in the symplectic picture}
One of the advantages of working with Gaussian states is that Gaussianity-preserving evolution corresponds to self-contained equations (\ref{eq:covariance_evolution_quadratic}) in the symplectic picture, by which we mean that the evolution of the covariance matrix and the vector of means can be traced without having to consider third- and higher-order correlation functions. As we show here, this property extends to eq. (\ref{eq:unitary_dissipator}), allowing one to study the evolution of quantum Gaussian states in the same fashion as in the case of Gaussian states.

Multiplying eq. (\ref{eq:unitary_dissipator}) by appropriate polynomials in the quadrature operators and taking the trace, we obtain the corresponding evolution of the first and second moments:
\begin{align} \label{eq:unitary_dissipator_moments}
\begin{split}
    \frac{d}{dt}\braket{\hat{\xi}_n\hat{\xi}_{n'}}=&
        \sum_{j=1}^M\gamma_j\braket{\hat{\xi}_{n,j}\hat{\xi}_{n',j}-\hat{\xi}_n\hat{\xi}_{n'}},\\
    \frac{d}{dt}\braket{\hat{\xi}_n}=&
        \sum_{j=1}^M\gamma_j\braket{\hat{\xi}_{n,j}-\hat{\xi}_n},
\end{split}
\end{align}
where
\begin{align} \label{eq:xi_transformed}
    \hat{\xi}_{n,j}\coloneqq \hat{U}_j^\dag\hat{\xi}_n\hat{U}_j,
\end{align}
denotes transformed quadrature operators.

Clearly, if the transformed quadrature operators are linear in the initial quadratures, then eqs (\ref{eq:unitary_dissipator_moments}) are closed with respect to the first two moments. In order for the new quadratures to have a physical meaning, they should also fulfill the canonical commutation relations. A generic transformation fulfilling these conditions is called a \emph{Bogoliubov transformation} \cite{Bogoliubov1958,Valatin1958,qft_2009}. In the case at hand, a generic Bogoliubov transformation reads explicitly
\begin{align} \label{eq:Bogoliubov_transformation}
    \hat{\xi}_{n,j} = \sum_{m=1}^{2N}(K_j)_{nm}\hat{\xi}_m,
\end{align}
where $K_j$ is a real symplectic matrix of size $2N\times 2N$. Under the assumption that the Lindbladians (\ref{eq:U_exponential_representation}) are generated by polynomials of at most second degree in quadrature operators, the matrices $K_j$ possess the convenient exponential representation (\ref{eq:K_exponential_representation}).

Taking the time derivative of the covariance matrix and the vector of means with eqs (\ref{eq:unitary_dissipator_moments}-\ref{eq:Bogoliubov_transformation}) at the input yields the symplectic picture equivalent to eq. (\ref{eq:unitary_dissipator}):
\begin{align} \label{eq:unitary_dissipator_Gaussian_evolution_general}
\begin{split}
    \frac{d}{dt}V=&
        \sum_{j=1}^M\gamma_j\left[K_j V K_j^T - V + F_j(\vec{\xi})\right],\\
    \frac{d}{dt}\vec{\xi}=&
        \sum_{j=1}^M\gamma_j K_j \vec{\xi},      
\end{split}
\end{align}
where
\begin{align}
\begin{split}
    F_j(\vec{\xi})=(K_j-\mathds{1})\vec{\xi}\vec{\xi}^T(K_j^T-\mathds{1}).
\end{split}
\end{align}

Note that in typical applications of the covariance matrix evolution, concerning e.g. quantum entanglement, the vector of mean values is irrelevant. For this reason, later on we will assume $\vec{\xi}(0)=0$, in which case $F_j(\vec{\xi})=0$ \footnote{To see this, consider the bottom line of eq. (\ref{eq:unitary_dissipator_Gaussian_evolution}). Solving for the vector of means, we immediately obtain $\vec{\xi}(t)=\exp[\Sigma_{j=1}^M\gamma_j K_j t]\vec{\xi}(0)$, which vanishes for $\vec{\xi}(0)=0$. Obviously, this implies $F_j(\vec{\xi})=0$.} and the evolution simplifies to
\begin{align} \label{eq:unitary_dissipator_Gaussian_evolution}
\begin{split}
    \frac{d}{dt}V=&
        \sum_{j=1}^M\gamma_j\left[K_j V K_j^T - V\right].
\end{split}
\end{align}
The corresponding explicit solutions are
\begin{align} \label{eq:V_t_superoperator_trick}
    V(t)=\sum_{j=0}^\infty p_j(t) K^j V(0) \big(K^T\big)^j
\end{align}
for a single Lindbladian and
\begin{align} \label{eq:V_t_superoperator_trick_general}
    V(t)=
        \sum_{k=0}^\infty \sum_{l_1\ldots l_j=1}^M p_{l_1\ldots l_j}(t) 
            K_{l_j}\ldots K_{l_1} V(0) 
            K_{l_1}^T\ldots K_{l_j}^T
\end{align}
for an arbitrary number of Lindbladians.

\vspace{0.5cm}

As we investigate below, depending on the number and nature of the unitary Lindblad operators, eq. (\ref{eq:unitary_dissipator}) and its symplectic representation (\ref{eq:unitary_dissipator_Gaussian_evolution}) can have radically different applications, ranging from random scattering to engineered dissipation.

\section{Random scattering}
\label{sec:classical_environment}
We now employ the collision model and kicked top dynamics to show that for a large number $M$ of non-commuting Lindblad operators, the discussed evolution constitutes a natural description of random scattering.

\subsection{Derivation from the collision model}
In the collision model \cite{collision_models_origin_Rau_1963,collision_models_review_Campbell_2021,collision_models_review_Ciccarello_2021}, the initial system is coupled to an infinite number of identical copies of ancilla $\hat{\eta}$. The total initial state is separable:
\begin{align} \label{eq:collision_model_total_initial}
    \hat{\rho}_T(0)=\hat{\rho}(0)\otimes\hat{\eta}\otimes\hat{\eta}\otimes\ldots
\end{align}
During the first time step $\Delta t$, a unitary operation $\hat{W}_1$ acts on the system and the first ancilla, after which the latter is traced out. The resulting state of the system is thus
\begin{align} \label{eq:collision_model_state_Delta_t}
    \hat{\rho}(\Delta t)=\Tr_{\eta}\{
        \hat{W}_1[\hat{\rho}(0)\otimes\hat{\eta}]\hat{W}_1^\dag\},
\end{align}
where $\Tr_{\eta}$ denotes partial trace over the ancilla. Since the corresponding total state has the same form as initially (\ref{eq:collision_model_total_initial}):
\begin{align}
    \hat{\rho}_T(\Delta t)=\hat{\rho}(\Delta t)\otimes\hat{\eta}\otimes\hat{\eta}\otimes\ldots
\end{align}
the second and further steps lead to analogous results as the first one. After $n$ steps
\begin{align}
    \hat{\rho}(n\Delta t)&=\Tr_{\eta}\{\hat{W}_n
        [\hat{\rho}[(n-1)\Delta t]\otimes\hat{\eta}]\hat{W}_n^\dag\}.
\end{align}
The unitaries $\hat{W}_n$ are typically assumed to have the elementary form \cite{collision_models_review_Ciccarello_2021}
\begin{align}
    \hat{W}_n = \exp\left[ -i (\hat{w}_S + \hat{w}_\eta + \hat{w}_{\textnormal{int},n} )\Delta t\right],
\end{align}
where the Hamiltonian $\hat{w}_S$ acts on the system, $\hat{w}_\eta$ acts on the bath, while $\hat{w}_{\textnormal{int},n}$ is responsible for interaction between the two. The last Hamiltonian may be step-dependent, while the others are assumed to be the same in each step. All three operators are time-independent.

Here, we employ a more general model \cite{collision_models_review_Campbell_2021}, in which the unitary operators are unrestricted. This gives the following general form
\begin{align} \label{eq:collision_model_W_general_integral}
    \hat{W}_n = \mathcal{T}\exp\left( -i \int_{(n-1)\Delta t}^{n \Delta t} d\tau\, \hat{w}_n(\tau)\right)
\end{align}
where $\mathcal{T}$ is the time-ordering operator and the time-dependent Hamiltonian $\hat{w}_n$ can act on both the system and the $n$-th ancilla in an arbitrary way. 

Clearly, by choosing the ancilla and the unitaries accordingly, we can use the collision model to emulate a wide range of dynamics. This fact, coupled with the relative conceptual simplicity, makes the collision model a popular tool in dealing with topics as varied as optics, thermodynamics and non-Markovianity, among others \cite{collision_models_review_Campbell_2021,collision_models_review_Ciccarello_2021}. Here, we use the collision model framework to derive the GKLS equation with $M$ unitary Lindblad operators.

For the ancillas, we choose qu$d$its of dimension $d=M+1$ in the ground state: 
\begin{align} \label{eq:collision_model_eta}
    \hat{\eta}=\ket{0}\bra{0}.
\end{align}
Furthermore, we choose unitary operations of the form
\begin{align} \label{eq:collision_model_W}
    \hat{W}_n = \left(\hat{\mathds{1}}\otimes\ket{0}\bra{0} +
        \sum_{j=1}^M\hat{U}_j\otimes\ket{j}\bra{j}\right)[\hat{\mathds{1}}\otimes\hat{O}(\Delta t)],
\end{align}
where $\hat{U}_j$ are arbitrary unitary operators with generators $\hat{h}_j$ [which, in the case of evolution preserving the convex hull of Gaussian states, are polynomials of at most second degree in quadrature operators] and $\hat{O}$ is a time-dependent unitary matrix defined by its action on the ancilla:
\begin{align} \label{eq:collision_model_O}
    \hat{O}(\Delta t)\ket{0} = \sqrt{1-\Delta t}\ket{0} 
        + \sqrt{\Delta t}\sum_{j=1}^M\sqrt{\gamma_j}\ket{j}.
\end{align}
As before, $\gamma_j\geqslant 0$ and $\sum_{j=1}^M\gamma_j=1$. With these inputs, eq. (\ref{eq:collision_model_state_Delta_t}) becomes
\begin{align}
    \hat{\rho}(\Delta t) = 
       \left[\hat{\mathds{1}} + \Delta t \sum_{j=1}^M \gamma_j \left(\hat{U}_j\cdot\hat{U}_j^\dag-\hat{\mathds{1}}\right)\right]
       \hat{\rho}(0),
\end{align}
Since in this setting we can easily recognize that $\hat{W}_n$ does not depend on the step number, each step corresponds to the same transformation. For $t=n\Delta t$ we therefore obtain
\begin{align}
    \hat{\rho}(t) = 
       \left[\hat{\mathds{1}} + \frac{t}{n} \sum_{j=1}^M \gamma_j \left(\hat{U}_j\cdot\hat{U}_j^\dag-\hat{\mathds{1}}\right)\right]^n
       \hat{\rho}(0).
\end{align}
In the continuous time limit $\Delta t \to 0$ taken simultaneously with $n\to\infty$, so that we approach a fixed value of time parameter $n\Delta t = t = \const$, we obtain the formal solution (\ref{eq:GKLS_formal_solution}) to the GKLS equation with generator (\ref{eq:unitary_Lindblad_equation_generator}). In other words, we recover the solution to the GKLS equation with $M$ unitary Lindbladians, as intended.

\subsection{Kicked top and scattering}
\label{sec:kicked_top_and_scattering}
We demonstrated that the GKLS evolution with unitary Lindbladians can be cast into the framework of collision models. To better understand implications of this fact, we now more deeply investigate the operator $\hat{W}_n$. As seen from eq. (\ref{eq:collision_model_W}), it is an unusual product of two sub-unitaries: a standard unitary operator and a time-independent ``kick''. 

Such structure is a staple in the kicked top model \cite{kicked_top_Haake_1978,kicked_top_Christensen_1979,kicked_top_Bhosale_2018}, defined by Hamiltonians of the form
\begin{align} \label{eq:Hamiltonian_kicked_top}
    \hat{H}_{\textnormal{kt}}(t) = \hat{H}_0(t) + \sum_m\delta(t-mT)\hat{V},
\end{align}
Here, the standard unitary dynamics generated by the base Hamiltonian $\hat{H}_0$ are periodically disturbed (with period length $T$) by the delta potential $\hat{V}$, leading to chaotic behaviour. Note that typically, the base Hamiltonian is assumed to be time-independent. However, the results remain qualitatively the same as long as the time-dependence of the Hamiltonian is well-behaved (i.e. not unbounded and discontinuous like the Dirac delta distribution). Due to its relative simplicity and ease of implementation in terms of qubits, the kicked top is the theoretical \cite{kicked_top_Bhosale_2018} and experimental \cite{kicked_top_experimental_Haake_2009} go-to model for testing the implications of dynamical chaos on quantum phenomena (such as, e.g., entanglement).

In Appendix \ref{app:kicked_top}, we show that the unitary operator (\ref{eq:collision_model_W}) can be obtained from the general eq. (\ref{eq:collision_model_W_general_integral}) by the kicked top Hamiltonian $\hat{w}_n=\hat{H}_{\textnormal{kt}}$ with
\begin{align} \label{eq:collision_model_kicked_top}
\begin{split}
    T=\Delta t, \quad \hat{H}_{0}(t) = \hat{o}_{n}(t), \quad \hat{V}=\sum_{j=1}^M\hat{h}_j\otimes\ket{j}\bra{j},
\end{split}
\end{align}
where $\hat{o}_n$ is the generator of $\hat{O}$ in the $n$-th step [see eq. (\ref{eq:U_O_generators}) in Appendix \ref{app:kicked_top} for definition]. Note that, because $\hat{w}_{n}$ acts only during the time interval $\big((n-1)\Delta t,n\Delta t\big]$, effectively only the $n$-th term in the sum (\ref{eq:Hamiltonian_kicked_top}) contributes.

Due to its close association with Poisson distribution \cite{unitary_lindbladians_Kummerer_1987,unitary_lindbladians_Frigerio_1989}, which describes random scattering through Poisson scatter theorem \cite{probability_book_Pitman}, the GKLS equation with unitary Lindbladians constitutes a valid model of random scattering. Our results make this interpretation explicit. 

Each collision can be seen as a single scattering event in the medium described by ancillas in the state (\ref{eq:collision_model_O}). Crucially, the probability that the system will be kicked by the $j$-th Hamiltonian $\hat{h}_j$ depends on $\gamma_j$ through eqs. (\ref{eq:collision_model_O}, \ref{eq:collision_model_kicked_top}). For a single Lindbladian, the system experiences identical scattering at every instant, quickly driving it towards a well-controlled stationary state (we investigate this in detail in the next section). However, as the number of mutually non-commuting unitaries entering the equation grows, so does the uncertainty in the outcome state. In particular, in the limit $M\to\infty$ the outcome probabilities $\gamma_j$ may be replaced by a probability measure $\mu(dU)$ on the unitary group, yielding a scattering integral \cite{unitary_lindbladians_kossakowski_1972}
\begin{align} \label{eq:scattering}
    \frac{d}{dt}\hat{\rho}=&
        \int d\mu(U)\left(\hat{U}\hat{\rho}\hat{U}^\dag-\hat{\rho}\right).
\end{align}
These results are consistent with previous findings \cite{unitary_lindbladians_Frigerio_1989,low_density_limit_Dumcke_1985} that unitary Lindbladians can be interpreted as the $S$-matrices of system interacting with a dilute gas.

The use of collision model is particularly appealing also when it comes to interpreting the role of the environment. Because the ancillas are traced out after each collision, during each step the system interacts with the same bath, fulfilling the expectation that the bath should not be influenced by the scattering (in particular, future scattering should not depend on previous events).

As a final remark, we note the the notion of quantum Gaussianity was born largely to address the fact that even though convex combination of Gaussian states are technically not Gaussian, i.e. they have non-Gaussian characteristic functions, they can be experimentally created and manipulated using the same methods as Gaussian states \cite{Gaussianity_resource_Takagi_2018,Gaussianity_resource_Albarelli_2018}. This makes Gaussian states and their convex combinations similar in practical applications. Our findings put the two families even closer, showing that states from the latter can be obtained from the former by simply subjecting them to random scattering, which may be regarded as pure noise. This result supports the developments made over the last decade to construct measures of quantum non-Gaussianity \cite{quantum_Gaussianity_Genoni_2013,quantum_Gaussianity_Hughes_2014,quantum_Gaussianity_Walschaers_2021}, which, contrary to measures of non-Gaussianity \cite{Gaussianity_measure_Genoni_2007,Gaussianity_measure_Genoni_2008,Gaussianity_measure_Genoni_2010,Gaussianity_measure_Ivan_2012,Gaussianity_measure_Mandilara_2012} do not assign positive values of the resource to convex combination of Gaussian states.

\section{Explicit solutions and stationary states}
\label{sec:explicit_solutions}
As seen, for a large number of unitary Lindblad operators, the considered evolution equation describes random scattering. However, the same equation equipped with a single Lindblad operator has well-controlled stationary states, as we proceed to show.

We start by deriving explicit solutions to the considered equation. Looking at eq. (\ref{eq:unitary_dissipator}), we can easily see that the stationary solutions $\hat{\rho}_\infty$ must commute with all the generators:
\begin{align} \label{eq:stationary_solutions_condition_standard}
\begin{split}
    0=[\hat{h}_j, \hat{\rho}_\infty] \quad \textnormal{for all }j.
\end{split}
\end{align}
As the number of non-commuting Lindbladians, and thus generators, approaches infinity, the evolution begins to describe pure decoherence, driving any initial state towards the maximally mixed state in the asymptotic time limit. This view was explored by us in the previous section. However, from the point of view of engineered dissipation, we expect only a few or even a single Lindbladian to appear, in which case it is possible to drive the system towards more useful stationary solutions.

Let us thus assume a single unitary Lindbladian generated by a hermitian operator $\hat{h}$ with eigendecomposition
\begin{align} \label{eq:H_eigendecomposition}
\begin{split}
    \hat{h}\ket{h_k}=h_k\ket{h_k},
\end{split}
\end{align}
where $h_k\in \mathbb{R}$ are assumed to be non-degenerate for simplicity. Since $\hat{h}$ is hermitian, its eigenvectors form a basis of the Hilbert space. In particular, one can write the initial density operator in this basis:
\begin{align} \label{eq:rho_in_H_basis}
\begin{split}
    \hat{\rho}=\sum_{k,k'}\rho_{kk'}^h\ket{h_k}\bra{h_{k'}}.
\end{split}
\end{align}  
Upon substituting into eq. (\ref{eq:unitary_dissipator}), we obtain
\begin{align}
\begin{split}
    \frac{d}{dt}\rho_{kk'}^h=\left(e^{-i(h_k-h_{k'})}-1\right)\rho_{kk'}^h.
\end{split}
\end{align}
This differential equation is easy to solve, yielding, after simplification, the general solution:
\begin{align} \label{eq:single_unitary_solution}
\begin{split}
    \rho_{kk'}^h(t)=e^{[\cos(h_k-h_{k'})-1]t}e^{-i[\sin(h_k-h_{k'})]t}\rho_{kk'}^h(0),
\end{split}
\end{align}
where $\rho_{kk'}^h(0)$ are the matrix elements of the initial state. 

The stationary states follow by taking the limit $t\to\infty$. All but the diagonal terms decay exponentially, leaving
\begin{align}
\begin{split}
    \hat{\rho}_\infty=\lim_{t\to\infty}\hat{\rho}(t)
        = \sum_{k} \lambda_k\ket{h_k}\bra{h_{k}}
\end{split}
\end{align}
with the final state's eigenvalues equal to
\begin{align}
\begin{split}
    \lambda_k = \rho^h_{kk}(0) = \bra{h_k}\hat{\rho}(0)\ket{h_k}.
\end{split}
\end{align}

A similar results holds in the symplectic picture. It is easy to show by using eqs (\ref{eq:symplectic_matrices_definition}, \ref{eq:K_exponential_representation}) that, since $J K_j=J e^{S_j J}=e^{JS_j} J$, eq. (\ref{eq:unitary_dissipator_Gaussian_evolution}) is equivalent to
\begin{align} \label{eq:JS_evolution}
\begin{split}
    \frac{d}{dt}(JV)=&
        \sum_{j}\left[e^{JS_j}(JV) e^{-JS_j} - JV\right].
\end{split}
\end{align}
Clearly, the stationary solutions $V_\infty$ are given by
\begin{align} \label{eq:stationary_solutions_condition_covariance}
\begin{split}
    0=[JS_j, JV_\infty] \quad \textnormal{for all }j.
\end{split}
\end{align} 
Therefore, just like in the standard picture the stationary solutions commute with the hermitian generators of evolution, in the symplectic picture the stationary solutions ``commute'' (commute after multiplication by $J$) with the symmetric generators of evolution.

Like before, let us consider a single unitary Lindbladian, which corresponds to a single symplectic operator. We denote the eigendecomposition of $JS$ by
\begin{align} \label{eq:JS_eigendecomposition_general}
\begin{split}
    JS \, \vec{s}_k = s_k \vec{s}_k.
\end{split}
\end{align}
Contrary to the hermitian generator $\hat{h}$ from the density operator picture, $JS$ does not have to be a normal matrix, meaning that its eigenvectors may not form a basis of the corresponding vector space. To solve the evolution equation explicitly, we consider the special case in which $JS$ is normal. This allows us to follow the reasoning from the density operator picture. 

We start by writing the initial covariance matrix as:
\begin{align} \label{eq:cv_in_JS_basis}
\begin{split}
    JV = \sum_{k,k'} (JV)_{kk'}^s \vec{s}_{k}\,\vec{s}_{k'}^\dag.
\end{split}
\end{align}  
Upon substituting into eq. (\ref{eq:JS_evolution}), we have
\begin{align}
\begin{split}
    \frac{d}{dt}(JV)_{kk'}^s=\left(e^{s_k-s_{k'}}-1\right)(JV)_{kk'}^s,
\end{split}
\end{align}
which is solved by
\begin{align} \label{eq:cv_solution}
\begin{split}
    (JV)_{kk'}^s(t)=\exp\left[\left(e^{s_k-s_{k'}}-1\right)t\right](JV)_{kk'}^s(0).
\end{split}
\end{align}
The asymptotic time limit depends on $s_k$. Denoting
\begin{align}
\begin{split}
    x_{kk'} &\coloneqq \re(s_k-s_{k'}), \\
    y_{kk'} &\coloneqq \im(s_k-s_{k'}), \\
    \zeta_{kk'} &\coloneqq \exp(x_{kk'})\cos(y_{kk'}),
\end{split}
\end{align}
we obtain, as $t\to\infty$,
\begin{align}\label{eq:stacases}
\begin{split}
    (JV)_{kk'}^s(t)\to\begin{cases}
    \infty & \zeta_{kk'} > 1, \\
    \exp[i\tan(y_{kk'}) t](JV)_{kk'}^s(0) & \zeta_{kk'} = 1, \\
    0 & \zeta_{kk'} < 1.
    \end{cases}
\end{split}
\end{align}

In the particular case of non-degenerate and purely imaginary $s_k$ (the latter happens whenever $K$ is passive, i.e. it is orthogonal in addition to being symplectic), the diagonal matrix elements approach the middle line (with $y_{kk}=0$), while the remaining elements approach zero. Consequently, the covariance matrix approaches the stationary solution
\begin{align}
\begin{split}
    JV_\infty = \lim_{t\to\infty}JV(t)
        = \sum_{k} \mu_k \vec{s}_{k}\,\vec{s}_{k}^\dag
\end{split}
\end{align}
with eigenvalues
\begin{align}
\begin{split}
    \mu_k = (JV)_{kk}^s(0) = \vec{s}_{k}^\dag JV(0)\vec{s}_{k}.
\end{split}
\end{align}
The corresponding symplectic eigenvalues can be then easily inferred from eq. (\ref{eq:mu_nu}).

On the other hand, for a generic choice of $K$, some matrix elements (\ref{eq:cv_solution}) diverge and some vanish exponentially with time. Thus, in this case, formally speaking there is no stationary solution to the equation considered. However, from a physical perspective, we focus on large rather than infinite times. From the point of view of the previous section, this may be interpreted as turning on the interaction with the environment for a given time, during which the system is subjected to a large, but finite number of infinitesimal kicks. Note that, in general, such kicks are not energy-preserving, since they may describe, e.g., squeezing transformations. In this regime, the covariance matrix becomes exponentially dominated by terms characterized in the first row of eq. (\ref{eq:stacases}). An example of such dynamics is investigated by us in the next section.

\section{Entanglement creation in two-mode states}
\label{sec:engineered_dissipation}
To illustrate the results derived in the previous section, we consider an engineered dissipation scenario, in which we use the discussed evolution equation for creation of two-mode entanglement from a system initially in the vacuum state
\begin{align} \label{eq:rho_vac}
    \hat{\rho}(0) = \ket{00}\bra{00},
\end{align}
which is separable and Gaussian. For the evolution, we choose a single Lindblad operator from the one-parameter family of unitary two-mode squeezing operators
\begin{align} \label{eq:squeezing_operator}
    \hat{L} = \hat{U}_r \coloneqq
        e^{i\hat{h}_r},
        \quad \hat{h}_r = -ir(\hat{a}_1^\dag\hat{a}_2^\dag-\hat{a}_1\hat{a}_2),
\end{align}
where $r>0$ is the squeezing strength and $\hat{a}_k\coloneqq\frac{1}{2}\left(\hat{x}_k+i\hat{p}_k\right)$ is the annihilation operator for mode $k$. 

Let us stress that, from the physical point of view, the evolution given by such a Lindblad operator is not at all equivalent to a ``smooth'' unitary evolution given by a squeezing Hamiltonian $\hat{H} = \hat{h}_r$. Instead, here, the squeezing should be understood as a series of regular, infinitely strong but infinitesimally short squeezing kicks, driving the system towards a high-energy state. In our case, the Hamiltonian behind the evolution is the kicked top Hamiltonian (\ref{eq:Hamiltonian_kicked_top}), with $\hat{h}_r$ entering at the level of the Dirac delta potential, as discussed in Section \ref{sec:kicked_top_and_scattering}.

We will proceed in two steps. First, we will certify that the evolved state is entangled. Then, we will quantify the amount of entanglement, showing that it is asymptotically unlimited.

\subsection{Certifying entanglement}
In the symplectic picture, the two-mode vacuum state is described by the covariance matrix
\begin{align} \label{eq:V_vac}
    V(0) = \frac{1}{2}\mathds{1}_4,
\end{align}
with $\vec{\xi}(0)=0$. As for the squeezing operator, it is well known \cite{De_Palma_2017} that 
\begin{align} \label{eq:squeezing_operator_Bogoliubov}
\begin{split}
    \hat{U}_{r}^\dag \hat{a}_1 \hat{U}_{r} = \cosh r \, \hat{a}_1 + \sinh r \, \hat{a}_2^\dag, \\
    \hat{U}_{r}^\dag \hat{a}_2 \hat{U}_{r} = \sinh r \, \hat{a}_1^\dag + \cosh r \, \hat{a}_2,    
\end{split}
\end{align}
Through eqs (\ref{eq:xi_transformed}-\ref{eq:Bogoliubov_transformation}), we can see that the above transformations corresponds to the symplectic matrix
\begin{align} \label{eq:K_entangling}
    K_r = \begin{bmatrix}
     \cosh r & 0 & \sinh r & 0 \\
     0 & \cosh r & 0 & -\sinh r \\
     \sinh r & 0 & \cosh r & 0 \\
     0 & -\sinh r & 0 & \cosh r \\
    \end{bmatrix}.
\end{align}
One can easily check that $K_r=\exp(JS_r)$ with
\begin{align} \label{eq:S_entangling}
    S_r = \begin{bmatrix}
     0 & 0 & 0 & r \\
     0 & 0 & r & 0 \\
     0 & r & 0 & 0 \\
     r & 0 & 0 & 0 \\
    \end{bmatrix}.
\end{align}
The matrix $JS_r$ is normal and has the following eigendecomposition [the notation is the same as in eq. (\ref{eq:JS_eigendecomposition_general})]:
\begin{align}
\begin{split}
    s_1 &= -r, \quad \vec{s}_1 = \frac{1}{\sqrt{2}}\left(0,1,0,1\right)^T, \\
    s_2 &= -r, \quad \vec{s}_2 = \frac{1}{\sqrt{2}}\left(-1,0,1,0\right)^T, \\ 
    s_3 &= r, \quad\;\;\: \vec{s}_3 = \frac{1}{\sqrt{2}}\left(0,-1,0,1\right)^T, \\
    s_4 &= r, \quad\;\;\: \vec{s}_4 = \frac{1}{\sqrt{2}}\left(1,0,1,0\right)^T. 
\end{split}
\end{align}
Using the methodology developed in the previous section, we can easily calculate the matrix $JV$ at any point in time. From the fact that $J^2=-\mathds{1}$, we then have $-J(JV)=V$, which explicitly reads
\begin{align} \label{eq:Vt}
\begin{split}
    V(t) &= \begin{bmatrix}
        A(t) & C(t) \\
        C(t) & A(t) \\
    \end{bmatrix}.
\end{split}
\end{align}
where 
\begin{align} \label{eq:ACt}
\begin{split}
    A(t) &= \frac{1}{2}e^{2t \sinh^2 r} \cosh \left(t\sinh 2r\right) \begin{bmatrix}
        1 & 0 \\
        0 & 1 \\
    \end{bmatrix}.\\
    C(t) &= \frac{1}{2}e^{2t \sinh^2 r} \sinh \left(t\sinh 2r\right) \begin{bmatrix}
        1 & 0 \\
        0 & -1 \\
    \end{bmatrix}.
\end{split}
\end{align}

Having obtained the time-evolved covariance matrix, we can use it to certify that the corresponding state is entangled. In the symplectic picture, a sufficient condition for the presence of entanglement in the system is given by the PPT criterion for continuous variable systems \cite{PPT,PPT_cv_systems}. The criterion states that, if the partial transposition of the state with respect to a given bipartition is not positive semi-definite, then the state is entangled with respect to this bipartition. In the case of two modes, the partially transposed state is not positive-semidefinite, and thus the state is entangled, if \cite{two-mode_gaussian_etc_proper_norm}
\begin{equation} \label{eq:PPT}
\begin{split}
    \tilde{\nu}_- < 1/2,
\end{split}
\end{equation}
where $\tilde{\nu}_-$ denotes the smallest symplectic eigenvalue of the covariance matrix of the partially transposed state:
\begin{align}
    V^{PT} = QVQ, \quad Q=\diag(1,1,1,-1).
\end{align}
Calculating the symplectic eigenvalues of $V^{PT}$ via the eigenvalues of $JV^{PT}$ as in eq. (\ref{eq:mu_nu}), we find that, in the case at hand,
\begin{align}
\begin{split}
    \tilde{\nu}_-(t) = \frac{1}{2}\exp\left[-\left({1-e^{-2r}}\right)t\right].
\end{split}
\end{align}
Evidently, the PPT criterion (\ref{eq:PPT}) for entanglement is fulfilled for all
\begin{align}
\begin{split}
    t > 0.
\end{split}
\end{align}
In other words, despite being initially separable, the state of the system is entangled throughout the whole evolution.

\subsection{Quantifying entanglement}
We certified that the considered dissipative evolution drives the, initially separable, system into an entangled state. We will now proceed to assess how much entanglement is contained in the time-evolved state. To this end, we consider a measure of entanglement called \emph{squashed entanglement} , one of the most prominent measures of entanglement \cite{squashed_entanglement_Matthias_2004,quantum_resource_theory_entanglement_HHHH_2009,distillation_entanglement_measures_Jeng_2019}. For a generic bipartite state $\hat{\sigma}_{AB}$, squashed entanglement is defined as
\begin{align}
\begin{split}
    \mathcal{E}_{\textnormal{sq}}(\hat{\sigma}_{AB}) \coloneqq \frac{1}{2}\inf_{\hat{\sigma}_{ABE}}
        I(A:B|E),
\end{split}
\end{align}
where $I(A:B|E)\coloneqq S_V(\hat{\sigma}_{AE})+S_V(\hat{\sigma}_{BE})-S_V(\hat{\sigma}_{E})-S_V(\hat{\sigma}_{ABE})$ is the conditional mutual information, $\hat{\sigma}_{X}$ are the (reduced) density operators of (sub)systems $X$ and the minimization is over all purifications $\hat{\sigma}_{ABE}$ of $\hat{\sigma}_{AB}$. Finally, 
\begin{align} \label{eq:von_Neumann_entropy}
\begin{split}
    S_V(\hat{\sigma})\coloneqq -\Tr \hat{\sigma}\ln\hat{\sigma}
\end{split}
\end{align} 
is the von Neumann entropy.

Like other entanglement measures defined in terms of minimization over some set of states, squashed entanglement is notoriously difficult to calculate, being an NP-hard computation problem \cite{squashed_entanglement_NP_hard_Huang_2014}. Here, we will not compute the squashed entanglement itself, but instead compute a lower bound for it and show that it is an asymptotically unbounded function of time.

We begin by observing that, due to the extremality of Gaussian states with respect to continuous, superadditive entanglement measures \cite{gaussian_extremality_Wolf_2006}, the squashed entanglement of any state $\hat{\sigma}$ is lower-bounded by the squashed entanglement of a Gaussian state $\hat{\sigma}_G$ with the same covariance matrix. Furthermore, squashed entanglement of any state is lower-bounded by so-called distillable entanglement $\mathcal{E}_{\textnormal{dist}}$ \cite{squashed_entanglement_Matthias_2004}, which, in turn, is lower-bounded by the coherent information \cite{coherent_information_Schumacher_1996,coherent_information_distillable_entanglement_bound_Devetak_2005} 
\begin{align}
\begin{split}
    I_{\mathcal{C}}(\hat{\sigma}) \coloneqq S_V(\hat{\sigma}_A) - S_V(\hat{\sigma}),
\end{split}
\end{align}
where $\hat{\sigma}_A=\Tr_B \hat{\sigma}$.

In our case, this means that we have the following chain of inequalities
\begin{align} \label{eq:E_sq_bound}
\begin{split}
    \mathcal{E}_{\textnormal{sq}}[\hat{\rho}(t)] & 
        \geqslant \mathcal{E}_{\textnormal{sq}}[\hat{\rho}_G(t)]
        \geqslant \mathcal{E}_{\textnormal{dist}}[\hat{\rho}_G(t)]\\
    & \geqslant I_{\mathcal{C}}[\hat{\rho}_G(t)] = S_V[\hat{\rho}_{G,A}(t)] - S_V[\hat{\rho}_G(t)],
\end{split}
\end{align}
where $\hat{\rho}_G(t)$ is a Gaussian state with the same covariance matrix (\ref{eq:Vt}) as our state and $\hat{\rho}_{G,A}(t)=\Tr_B \hat{\rho}_G(t)$. Crucially, both von Neumann entropies on the r.h.s. are simple functions of the symplectic eigenvalues of the respective state. Let us define the auxiliary function
\begin{align}
\begin{split}
    f(x) \coloneqq (x+1/2)\ln(x+1/2)-(x-1/2)\ln(x-1/2).
\end{split}
\end{align}
Then, for a one- or two-mode Gaussian state $\hat{\sigma}_G$ with covariance matrix $V_{\hat{\sigma}}$ \cite{Gaussian_von_Neumann_entropy_Serafini_2003}
\begin{align}
\begin{split}
    S_V(\hat{\sigma}_G) = \sum_{j=1}^N f\left[\nu_j(V_{\hat{\sigma}})\right],
\end{split}
\end{align}
where $N$ is the number of modes. In the case at hand, it is easy to calculate that the symplectic eigenvalues of the covariance matrix (\ref{eq:Vt}) equal
\begin{align}
\begin{split}
    \nu_1(t) = \nu_2(t) = \frac{1}{2}e^{2t\sinh^2 r}\equiv \nu(t).
\end{split}
\end{align}
On the other hand, one can easily see from the definition (\ref{eq:covariance_matrix}) that the reduced covariance matrix $V_A$ corresponding to the first mode is given by the upper-left block of (\ref{eq:Vt}), i.e. $V_A = A(t)$. The only symplectic eigenvalue of $V_A$ equals
\begin{align}
\begin{split}
    \nu_A(t) = \frac{1}{2}\cosh(t\sinh 2r)e^{2t\sinh^2 r}.
\end{split}
\end{align}
Using the last four equations in eq. (\ref{eq:E_sq_bound}), we finally obtain
\begin{align}
\begin{split}
    \mathcal{E}_{sq}[\hat{\rho}(t)] \geqslant I_{\mathcal{C}}[\hat{\rho}_G(t)] 
        = f[\nu_A(t)] - 2f[\nu(t)].
\end{split}
\end{align}
The above lower bound for squashed entanglement, and therefore also squashed entanglement itself, grows indefinitely. To show this, we first calculate that
\begin{align}
\begin{split}
    I_{\mathcal{C}}[\hat{\rho}_G(t)] = e^{2z}\ln\tanh z 
        + \ln\frac{2\left(e^{2z}\cosh[2\coth(r)z]+1\right)}{e^{4z}-1}.
\end{split}
\end{align}
where we denoted $z\coloneqq t \sinh^2 r$ for shortness. For very large $t$, corresponding to very large $z$, the first term on the r.h.s. approaches the constant value of $-2$. In the second term, $\cosh[2\coth(r)z]$ approaches $e^{2\coth(r)z}/2$, which means that the logarithm behaves like $\ln e^{4z[\coth(r)-1]}=4z[\coth(r)-1]$. It follows that
\begin{align}
\begin{split}
    I_{\mathcal{C}}[\hat{\rho}_G(t)] \xrightarrow[t \to \infty]{}
        &-2 + 4z[\coth(r)-1] \\
        =& -2 + 4 t \sinh^2(r)[\coth(r)-1],
\end{split}
\end{align}
where, to be explicit, in the bottom line we went back to the parametrization in terms of $t$. Clearly, this is a linear function in $t$ with positive slope, since $\coth(r)>1$ for all $r>0$. Therefore, $I_{\mathcal{C}}[\hat{\rho}_G(t)]$ is asymptotically infinite, and thus the same is also true for squashed entanglement itself. This is what we wanted to show.

\section{Concluding remarks}
\label{sec:conclusion} 
Motivated by recent findings in resource theories of non-Gaussianity, we developed a model of dissipative evolution which preserves the set of quantum Gaussian states without preserving the set of Gaussian states itself. We showed that, while such a model constitutes a natural description of random scattering, it can also be applied to engineered dissipation, as showcased through an example of entanglement creation in two-mode states. Finally, the model is fully compatible with the symplectic (covariance matrix) picture of quantum states, allowing one to study it with the same tools that are typically used for Gaussian states.

Besides applications to phenomena that include random scattering, as well as engineered dissipation, our findings suggest the following directions for future research. To start with, let us briefly denote the generator of Gaussian evolution (\ref{eq:covariance_evolution_quadratic}) by $\mathcal{L}_G$ and the generator of the evolution (\ref{eq:unitary_dissipator}) based on unitary Lindbladians by $\mathcal{L}_{cG}$. Because both $\mathcal{L}_G$ and $\mathcal{L}_{cG}$ preserve the set of quantum Gaussian states, then, by Trotter's formula \cite{Trotter}
\begin{align} \label{eq:Trotter}
\begin{split}
    e^{(\mathcal{L}_{G} + \mathcal{L}_{cG})t} = \lim_{n\to\infty} 
        \left(e^{\mathcal{L}_{G} t/n} e^{\mathcal{L}_{cG} t/n}\right)^n
\end{split}
\end{align}
the combined generator $\mathcal{L}_G + \mathcal{L}_{cG}$ also does. Therefore, from the point of view of dynamics of quantum Gaussian states, the discussed generator can be seen not only as an alternative to the Gaussian model, but also as its extension. For example, it could be used to introduce generic quantum Gaussian noise, especially in the form of the scattering integral (\ref{eq:scattering}), into an otherwise Gaussian system. 

Furthermore, while operations preserving the set of Gaussian states are fully characterized \cite{Gaussian_operators_characterization_De_Palma_2015}, an analogous problem was not resolved for quantum Gaussian states, partially due to the lack of one-to-one correspondence with the set of states with positive Wigner distribution \cite{convex_Gaussian_states_Wigner_positivity_Brocker_1995}. This leads to the following question: what other evolution models preserve the set of quantum Gaussian states but not the set of Gaussian states? What physical scenarios can they describe? An immediate generalization of our results would be to replace $\mathcal{L}_{cG}$ by $\mathcal{L}\cdot =\sum_{k}\left(\theta_k\cdot - \hat{\mathds{1}}\right)$, with $\theta_k$ being arbitrary Gaussian channels. One can easily check that such generator preserves the set of quantum Gaussian states. It would be interesting to see if this is the most general generator with this property, and if not, how it could be generalized further.

\acknowledgements 
We thank the anonymous referees for several helpful comments and remarks. Tomasz Linowski and {\L}ukasz Rudnicki acknowledge support by the Foundation for Polish Science (IRAP project, ICTQT, contract no. 2018/MAB/5, co-financed by EU within Smart Growth Operational Programme).

\bibliography{report}{}
\bibliographystyle{obib}

\appendix

\section{Rewriting the operator (\ref{eq:collision_model_W}) in terms of a kicked top Hamiltonian} 
\label{app:kicked_top}
\setcounter{equation}{0}
\renewcommand{\theequation}{\ref{app:kicked_top}\arabic{equation}}
In this appendix, we show that the unitary operator (\ref{eq:collision_model_W}) can be obtained by substituting the kicked top Hamiltonian $\hat{w}_n=\hat{H}_{\textnormal{kt}}$ with inputs (\ref{eq:collision_model_kicked_top}) into eq. (\ref{eq:collision_model_W_general_integral}). In other words, we show that the operators
\begin{align}
    \hat{X}_n &= \left(\hat{\mathds{1}}\otimes\ket{0}\bra{0} +
        \sum_{j=1}^M\hat{U}_j\otimes\ket{j}\bra{j}\right)\hat{O}(\Delta t),
        \label{eq:X_n}\\
    \hat{Y}_n &= \mathcal{T}\exp\left( -i \int_{(n-1)\Delta t}^{n \Delta t} d\tau\, \hat{w}_n(\tau)\right),
\end{align}
are identical for
\begin{align}
    \hat{w}_{n}(\tau) = \hat{o}_n(\tau) + \delta(\tau-n\Delta t)\sum_{j=1}^M\hat{h}_j\otimes\ket{j}\bra{j},
\end{align}
with $\hat{U}_j = e^{-i \hat{h}_j}$ and
\begin{align} \label{eq:U_O_generators}
    \hat{O}(\Delta t) = \mathcal{T}\exp\left( -i \int_{(n-1)\Delta t}^{n \Delta t} d\tau\, \hat{o}_n(\tau)\right).
\end{align}

We begin by observing that $\hat{Y}_n$ can be recast into
\begin{align}
\begin{split}
    \hat{Y}_n =
        \lim_{\epsilon\to 0}&\:\mathcal{T}\exp
        \left(-i\int_{n\Delta t-\epsilon/2}^{n \Delta t+\epsilon/2} d\tau\, \hat{w}_n(\tau)\right)\\
        &\times\mathcal{T}\exp
        \left(-i\int_{(n-1)\Delta t}^{n \Delta t-\epsilon/2} d\tau\, \hat{w}_n(\tau)\right).
\end{split}
\end{align}
Provided $\hat{o}_n$ is a well-behaved function of time [which can be inferred from the well-behaved nature of its exponential (\ref{eq:collision_model_O})], its contribution to the first integral vanishes in the limit. At the same time, the delta distribution integrates to one. See e.g. \cite{delta_treatment_Blume_1978} for rigorous treatment. In conclusion,
\begin{align}
\begin{split}
    \mathcal{T}\exp
        \left(-i\int_{n\Delta t-\epsilon/2}^{n \Delta t+\epsilon/2} d\tau\, \hat{w}_n(\tau)\right)
        \to e^{-i\sum_{j=1}^M\hat{h}_j\otimes\ket{j}\bra{j}}.
\end{split}
\end{align}
In the second integral, the situation is reversed. Because the integral does not contain the point $\tau = n\Delta t$, the delta distribution does not contribute and we can simply put $\hat{w}_n = \hat{o}_n$. Thus, based on eq. (\ref{eq:U_O_generators}),
\begin{align}
\begin{split}
    \mathcal{T}\exp
        \left(-i\int_{(n-1)\Delta t}^{n \Delta t-\epsilon/2} d\tau\, \hat{w}_n(\tau)\right)
        \to \hat{O}(\Delta t).
\end{split}
\end{align}
Combining the last three expressions, we obtain
\begin{align} \label{eq:Y_n}
\begin{split}
    \hat{Y}_n = e^{-i\sum_{j=1}^M\hat{h}_j\otimes\ket{j}\bra{j}} \hat{O}(\Delta t).
\end{split}
\end{align}
Because the generator of the exponential on the r.h.s. is diagonal in the second subsystem's number basis, the exponentiation can be explicitly performed, quickly yielding
\begin{align}
\begin{split}
    e^{-i\sum_{j=1}^M\hat{h}_j\otimes\ket{j}\bra{j}} &= \hat{\mathds{1}}\otimes\ket{0}\bra{0} +
        \sum_{j=1}^M\hat{U}_j\otimes\ket{j}\bra{j}.
\end{split}
\end{align}
Clearly, this makes eq. (\ref{eq:Y_n}) identical to (\ref{eq:X_n}), which is what we wanted to prove.

\end{document}